\documentclass[aps,prl,twocolumn,amsmath,amssymb]{revtex4} 
\usepackage[english]{babel}
\usepackage{latexsym}
\usepackage{graphics}
\usepackage{subfigure}
\usepackage{epsfig}
\begin{document}

\title{Atomic Bose-Fermi mixtures in an optical lattice}
\author{M. Lewenstein$^{1}$, L. Santos$^{1}$, M. A. Baranov$^{1,2}$, and H. Fehrmann$^{1}$}  
\address{(1)Institut f\"ur Theoretische Physik, Universit\"at Hannover, 
D-30167 Hannover,Germany}
\address{(2)RRC, Kurchatov Institute, Kurchatov square 1, 123182 Moscow, Russia}
\begin{abstract}
A mixture of ultracold bosons and fermions placed 
in an optical lattice constitutes a novel kind of quantum gas, and leads
to phenomena, which so far have been discussed neither 
in atomic physics, nor in condensed matter physics.  We
discuss the phase diagram at low temperatures, and in
the limit of strong atom-atom interactions, and predict the existence of
quantum  phases that involve pairing of fermions with one or more
bosons, or, respectively, bosonic holes. The resulting composite fermions
may form, depending on the system parameters, a normal Fermi liquid, a
density wave, a superfluid liquid, or an insulator
with fermionic domains.  We discuss the feasibility for
observing such phases in current experiments.
\end{abstract}
\maketitle


Since the first observation of Bose-Einstein condensation 
(BEC) in atomic vapors \cite{BEC}, atomic physics has been
constantly approaching research topics traditionally associated with
condensed matter physics, such as the analysis of superfluidity in BEC
\cite{Superfl}, 
or the on-going intensive
search for the Bardeen-Cooper-Schrieffer (BCS) superfluid transition in
ultracold atomic Fermi gases
\cite{FFexperim,BFexperim}. Recently, 
striking experimental developments have driven a rapidly-growing
interest on strongly correlated systems in atomic physics. 
In this sense, the control of the interatomic
interactions via Feshbach resonances
\cite{Feshbach} has become particularly interesting, opening the 
way towards strongly-interacting gases \cite{jila}. 
Additionally, strong correlations are predicted to play a
dominant role in low-dimensional systems, such as 1D Bose gases
(Tonks-Girardeau regime) \cite{tonks}, 1D Fermi systems
(Luttinger liquid) \cite{Recati}, or rapidly-rotating 2D gases,
where the physics resembles that of fractional quantum
Hall effect \cite{belen}.


The recent observation of the superfluid (SF) to Mott Insulator 
(MI) transition in ultracold atoms in optical lattices \cite{bloch},
predicted in Ref.~\cite{jaksch}, constitutes up to now the most
spectacular example of phenomena related to strongly-correlated
atomic gases. In this experiment (performed with $^{87}$Rb atoms), 
by changing the laser intensity and/or detuning, one can control 
the tunneling to neighboring sites as well as the strength of the on-site 
repulsive interactions, and therefore one is able to switch between the SF phase 
(dominated by the tunneling) and the MI phase with a fixed number of atoms 
per site \cite{bloch}. 


The possibility of sympathetic cooling of fermions with bosons 
has lead to several recent experiments on 
trapped ultracold Bose-Fermi mixtures \cite{BFexperim}. 
So far, temperatures $T\sim 0.05$ $T_F$ have been obtained, 
where $T_F$ is the Fermi temperature at which the
Fermi gas starts to exhibit quantum degeneracy (typically of the 
order of 10 $\mu$K). 
Although the main goal of these 
experiments is to achieve the BCS transition in atomic Fermi
gases, several groups recently show a growing interest
towards the physics of ultracold Bose-Fermi mixtures themselves, 
including the analysis of the ground-state properties, stability, 
excitations, and the effective Fermi-Fermi interaction 
mediated by the bosons \cite{BFmixt}. Additionally, 
new experimental developments 
have attracted the attention towards the behavior of these mixtures in
1D geometries \cite{BF1D}, 
and optical lattices \cite{Eisert,BFLatt}.


In this letter, we investigate a Bose-Fermi lattice gas, 
i.e. a mixture of ultracold bosonic and fermionic atoms in an
optical lattice. This system is somewhat similar to the Bose lattice gas
of Ref. \cite{bloch}, yet much more complex and with a richer behavior at
low temperatures. 
We discuss the limit of strong 
atom-atom interactions (strong
coupling regime) at low temperatures. 
Our main prediction concerns the existence of novel
quantum phases that involve pairing of fermions with
one or more bosons, or, respectively, bosonic holes, depending on
the sign of the interaction between fermions and bosons \cite{Kaganfermibose}. 
The resulting composite fermions may form a normal Fermi liquid, 
a density wave, a superfluid, or an insulator with fermionic domains,
depending on the parameters characterizing the system. At the end 
of this paper, we discuss the experimental feasibility of the 
predicted phases.


The lattice potential is
practically the same for both species in a 
$^7$Li-$^6$Li mixture, and accidentally
very similar for the $^{87}$Rb-$^{40}$K case 
(for detunings corresponding to the wavelength 
$1064$ nm of a Nd:Yag laser \cite{Weidemuller}). Due to the
periodicity of this potential, energy bands are formed. 
If the temperature is low enough and/or the
lattice wells are sufficiently deep, the atoms 
occupy only the lowest band. Of course, for fermions this is
only possible if their number is smaller than the number of
lattice sites (filling factor $ \rho_F\le 1$). To describe the system
under these conditions, we choose a particularly suitable set of single
particle states in the lowest band, the so-called Wannier states,
which are localized at each lattice site. The system is then
described by the tight-binding Bose-Fermi-Hubbard (BFH) model 
(for a derivation from a microscopic model, see Ref.~\cite{Eisert}), 
which is a generalization of the fermionic Hubbard model, 
extensively studied in condensed matter theory (c.f. \cite {Auerbach}):
\begin{eqnarray}
H_{\mathrm{BFH}}&=&-\sum_{\left\langle ij\right\rangle }
(J_B b_{i}^{\dagger }b_{j}+J_F f_{i}^{\dagger }f_{j}+ {\rm h.c.}) \nonumber \\
&+&\sum_{i}\left[
\frac{1}{2}Vn_{i}(n_{i}-1)-\mu n_{i}\right] +U\sum_{i}n_{i}m_{i}, 
\label{Hamiltonian} 
\end{eqnarray}
 where $b_{i}^{\dagger }$, $b_{j}$,
$f_{i}^{\dagger }$, $f_{j}$ are the bosonic and fermionic
creation-annihilation operators, respectively, $n_{i}=b_{i}^{\dagger
}b_{i}$, $m_{i}=f_{i}^{\dagger }f_{i}$, and $\mu $ is the bosonic
chemical potential. The fermionic chemical potential is absent in 
$H_{\mathrm{BFH}}$, since the fermion number is fixed.
The BFH model
describes: i) nearest neighbor boson (fermion) hopping, with an  
associated negative energy, $ -J_{B}$ ($-J_{F}$); 
in the following we assume $J_{F}=J_{B}=J$, while the more general case 
of different tunneling rates will be analyzed elsewhere; 
ii) on-site repulsive boson-boson interactions with
an associated energy $V$; iii) on-site boson-fermion interactions with an
associated energy $U$, which is positive (negative) for repulsive
(attractive) interactions. 

In this letter we are interested in the strong-coupling limit, 
$J\ll U,V$, which can be easily reached experimentally by increasing the 
lattice intensity.
We first analyze the case of vanishing hopping ($J=0$). For 
the simplest case of $U=0$, at
zero temperature, the fermions can occupy any available many-body state,
since the energy of the system does not depend on their configuration. 
Bosons on the contrary are necessarily in the MI state with exactly 
$\tilde n=[\tilde\mu]+1$ bosons per
site, where $[\tilde\mu]$ denotes the integer part of $\tilde\mu=\mu /V$. 
For small $
|U|\neq 0$, the system is only perturbatively affected. However, if
$ U>0 $ is sufficiently large, $U>\mu-(\tilde n-1)V$, the fermions push the
bosons out of the sites that they occupy. Hence, 
localized composite fermions are formed, consisting of one
fermion and the corresponding number of missing bosons (bosonic holes). 
Similarly, if $U<\mu-\tilde nV$, 
the fermions will attract bosons to their sites, and again
localized composite fermions are formed, but now consisting of
one fermion and the corresponding number of bosons. 

\begin{figure}[ht] 
\begin{center}
\psfig{file=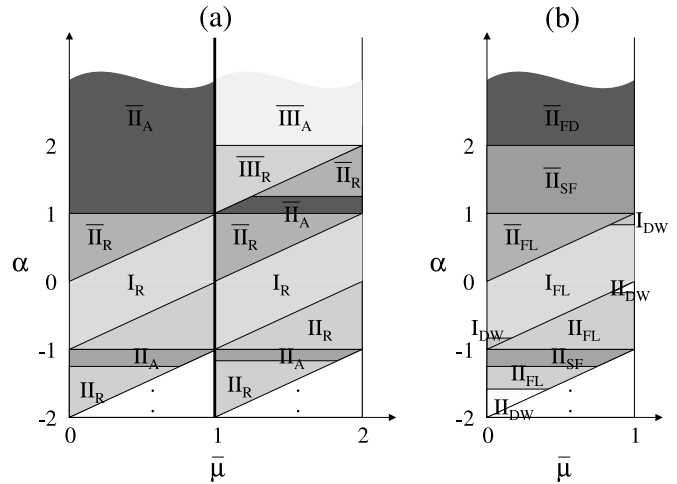,width=6.8cm,angle=270}
\end{center} 
\vspace*{-0.5cm}
\caption{(a) Phase space as a function of $\bar\mu$ and $\alpha=U/V$. See text for details on the 
notation; (b) Full phase diagram for the region $0<\bar\mu<1$, 
for $\rho_f=0.4$ and $J/V=0.02$. 
Different phases are present, including fermionic domains (FD), 
superfluid (SF), Fermi liquid (FL) and density-wave phase (DW).}
\label{fig:2}  
\end{figure}

Fig.~\ref{fig:2}(a) shows the phase diagram of the system in the $
\alpha -\bar{\mu}$ plane, where $\alpha =U/V$.
Quite generally, for $\bar{\mu}-[\bar{\mu}]+s>\alpha
>\bar{\mu}-[\bar{\mu}]+s-1$, we obtain that 
$s$ holes (or, for $s<0$, $-s$ bosons) form with a single fermion
a composite fermion, annihilated by 
$
\tilde f_i=\sqrt{(\tilde n-s)!/\tilde n!}(b_i^\dag)^s f_i
$
(
$
\sqrt{\tilde n!/(\tilde n-s)!}(b_i)^{-s} f_i
$
). Since the maximal number of holes is
limited by $\tilde n$, $s$ must not be greater than $\tilde n$; it can,
on the other hand, attain arbitrary negative integer values, i.e., 
we may have fermion composites of one fermion and many bosons in
the case of very strong attractive interactions, $\alpha <0$, and
$|\alpha |\gg 1$. In Fig.~\ref{fig:2}(a) the different regions in the phase
diagram are denoted with Roman numbers I, II, III, IV etc, which denote 
the number of particles that form the corresponding
composite fermion. Additionally, a bar over a Roman number indicates
composite fermions formed by one bare fermion and bosonic holes, rather
than bosons. 

Although our composite fermions neither move, nor
interact with each other ($J=0$), 
the phase diagram is quite complex. As a result, switching on a small, but
finite hopping, leads to an amazingly rich physics.
The latter can be investigated on the basis of an effective theory 
for composite fermions, which can be derived using degenerate 
perturbation theory (to second order in $J$) along the lines of the derivation 
of the $t-J$ model (see e.g. Ref.~\cite {Auerbach}). 
Remarkably, the resulting effective model is universal
for all the distinct regions in the phase diagram in Fig.~\ref{fig:2}(a),
and the corresponding Hamiltonian
\begin{equation}
H_{eff}=-J_{eff}\sum_{\left\langle ij\right\rangle }(\tilde f_{i}^{\dagger
}\tilde f_{j} +{\rm h.c.} )+K_{eff}\sum_{\left\langle ij\right\rangle }\tilde m_{i}
\tilde m_{j}
\label{EffHamiltonian}
\end{equation}
is determined by two effective parameters describing: 
i) nearest neighbor hopping of composite fermions with the corresponding  
negative energy $-J_{eff}$; ii) nearest neighbor
composite fermion-fermion interactions with the associated energy
$K_{eff}$, which may be repulsive ($>0$) or attractive ($<0$). In 
Eq.~(\ref{EffHamiltonian}) 
we employ the number operator $\tilde m_i=\tilde f_i^\dagger \tilde f_i$. 
This effective model is equivalent to that of spinless
interacting fermions (c.f. \cite{Sachdev,Shankar}), and, despite its 
simplicity, has a rich phase diagram. 
The coefficient $K_{eff}$ has the universal form 
\begin{eqnarray}
&&K_{eff}=2\frac{J^2}{V}
\left\{
\frac{\tilde n(\tilde n+1-s)}{1+\alpha-s}+\frac{(\tilde n-s)(\tilde n+1)}{1-\alpha+s}
\right\delimiter 0 \nonumber \\
&&\left\delimiter 0
+\frac{1}{\alpha s}-\tilde n(\tilde n+1)-(\tilde n-s)(\tilde n +1-s)
\right \},
\end{eqnarray}
whereas the dependence of $J_{eff}$ on $J$, $V$, and $U$, has 
different forms in different regions of Fig.~\ref{fig:2}(a). 
E.g. for $0< \tilde\mu <1$, $J_{eff}=J$ (in $I$), 
$2J^2/\alpha V$ (in $\overline{II}$), $4J^2/|\alpha|V$ (in $II$), etc.
The physics of the effective model 
is determined by the ratio $\Delta =K_{eff}/2J_{eff}$,
and by the sign of $K_{eff}$. In Fig.~\ref{fig:2}(a) the 
subindex $A$ ($R$) denotes attractive (repulsive) interactions.

The problem of finding the ground state of
the BFH model is then reduced to the analysis of the ground state of 
the spinless Fermi model (\ref{EffHamiltonian}). 
In the case of a repulsive effective interaction, $%
K_{eff}>0$, and filling fraction close to zero, $\rho _{F}\ll 1$, or one, $%
1-\rho _{F}\ll 1$, the ground state of $H_{eff}$ corresponds to a Fermi
liquid (a metal), and is well described in the Bloch representation. In the
considered cases, the relevant momenta are small compared to the inverse
lattice constant (the size of the Brillouin zone). One can thus take the
continuous limit, in which the first term in $H_{eff}$ corresponds to a
quadratic dispersion with a positive (negative) effective mass for particles
(holes), while the second term describes $p$-wave interactions. 
The lattice is irrelevant in this limit, and the system is
equivalent to a Fermi gas of spinless fermions (for $\rho _{F}\ll 1$), or
holes (for $1-\rho _{F}\ll 1$). Remarkably, this gas is weakly interacting
for every value of $K_{eff}$, even when $K_{eff}\rightarrow \infty $. The
latter case corresponds to the exclusion of the sites that surround an
occupied site from the space available for other fermions. As a result, the
scattering length remains finite, being of the order of the lattice spacing.
Therefore, $1-\rho _{F}$ ($\rho _{F}$) acts as the gas parameter for the gas
of holes (particles). This picture can be rigorously justified using renormalization
group approach \cite{Shankar}.

The weakly-interacting picture becomes inadequate near half-filling, $\rho
_{F}\rightarrow 1/2$, and for large $\Delta $, where the effects of the
interactions between fermions become important, and one expects the
appearance of localized phases. A physical insight on the properties of this
regime can be obtained by using Gutzwiller ansatz (GA) \cite{Gutzwiller}, 
in which the ground
state is a product of on-site states with $0$ or $1$ composites,  
$\prod_i (\cos\theta_i/2 |1\rangle_i+\sin\theta_i/2 e^{\phi_i}|0\rangle_i)$, 
and which is in fact well-suited for
describing the states with reduced mobility and, therefore, with small
correlations between different sites. Such an approach allows to determine
the boundaries of various quantum phases relatively well in 3D, 2D, and even
1D, but does not provide the correct description of correlations and
excitations; these failures become particularly important in 1D, 
where, strictly speaking, the GA approach is inappropriate. For $%
K_{eff}>0$ the GA approach maps $H_{eff}$ onto the classical
antiferromagnetic spin model with spins of length $1$, 
$\vec S_i=(\sin\theta_i\cos\phi_i,\sin\theta_i\sin\phi_i,cos\theta_i)$ 
\cite{Auerbach}. The
corresponding ground state is a spin-flop (canted) antiferromagnet \cite
{Auerbach,Sachdev} with a constant density, provided $\Delta <\Delta
_{crit}=(1+m_{z}^{2})/(1-m_{z}^{2})$, where the ''magnetization per spin''
is $m_{z}=2\rho _{F}-1$. When $\Delta >\Delta _{crit}$, the GA ground state
of the classical spin model exhibits modulations of $m_{z}$ with a
periodicity of two lattice constants. We expect that the employed GA
formalism predicts the phase boundary $\Delta _{crit}$ accurately for $\rho
_{F}$ close to $1/2$. Coming back to the composite fermion picture, we
predict thus that the ground state for $\Delta <\Delta _{crit}$ 
is a Fermi liquid, while for $\Delta >\Delta_{crit}$ it is a 
density wave.
For the special case of half filling, $\rho _{F}=1/2$, the ground state is
the so-called checkerboard state, with every second site occupied by one
composite fermion. One should stress that the GA value of $\Delta _{crit}$
is incorrect for filling factors $\rho _{F}$ close to $0$ or $1$. In
particular, the GA approach predicts that $\Delta _{crit}$ tends gradually
to infinity and the density wave phase gradually shrinks as $\rho
_{F}\rightarrow 0$ or $1$, i.e. $1-m_{z}^{2}\rightarrow 0$. As discussed
above, an analysis beyond the GA approach shows the disappearance of the
density wave phase already for a finite non-zero value of $1-m_{z}^{2}$.

The situation is different when the effective interaction is attractive, $%
K_{eff}<0$, which in the spin description corresponds to ferromagnetic spin
couplings. In the GA approach the ground state for $0>\Delta \geq -1$ is
ferromagnetic and homogeneous. In this description, 
fixing the fermion number means fixing the $z$ component 
of the magnetization $M_{z}=N(2\rho
_{F}-1)$. When $|\Delta |\ll 1$, and $\rho _{F}$ is close to zero (one),
i.e. low (high) lattice filling, a very good approach to the 
ground state is given by a BCS ansatz \cite{DeGennes},
in which the composite fermions (holes) of opposite momentum build $p$-wave
Cooper pairs, $\prod_{\vec k}(v_{\vec k}|00\rangle_{\vec k,-\vec k}+
u_{\vec k} |1,1\rangle_{\vec k,-\vec k})$, where $v_{\vec k}$ and 
$u_{\vec k}$ are the coefficients of the Bogoliubov transformation.
The ground state becomes more complex for arbitrary $\rho _{F}$%
, and for $\Delta $ approaching $-1$ from above. The system becomes strongly
correlated, and the composite fermions in the SF phase may build not only
pairs, but also triples, quadruples etc. The situation becomes simpler when $%
\Delta <-1$. In the spin picture the spins form then ferromagnetic domains
with spins ordered along the $z$-axis. In the fermionic language 
this corresponds to the formation of domains of composite fermions (''domain'' insulator). 
This mean-field result is in fact exact.

Fig.~\ref{fig:2}(b) shows the cases $I$, $II$ and 
$\overline{II}$ for $0<\mu<1$, with the predicted quantum phases.
The observation of the predicted phases, 
constitutes a challenging, but definitely accessible, 
goal for experiments. Systems of different dimensionalities 
are nowadays achievable by controlling the 
potential strength in different directions \cite{lowD}.
The conditions for the exclusive occupancy of the 
lowest band, and for $J\ll V,U$, are fulfilled for 
sufficiently strong lattice potentials, as those typically employed 
in current experiments \cite{bloch} ($10$-$20$ recoil energies). 
Additionally, our $T=0$ analysis is valid 
for $T$ much lower than the smallest
energy scale in our problem, namely the tunneling rate. 
This regime is definitely
accessible for sufficiently large interactions.
In typical experiments, the presence of an inhomogeneous trapping potential 
leads to the appearance of regions of different phases \
\cite{jaksch,Muramatsu}, and it is crucial for the observation of MI phases \cite{bloch}. 
The inhomogeneity controls thus the bosonic chemical potential, which  
can also be tailored by changing the number
of bosons in the lattice, regulating the strength of the lattice
potential, and/or modifying the interatomic interactions by means of
Feshbach resonances \cite{Feshbach}.
We would like also to note that 
for $J\ll V$, phases $I$, $II$ and $\overline{II}$ are easier to study, since 
the fermions, or composite fermions,  
attain effective hopping energies that
are not too small, and can compete with the effective
interactions $K_{eff}$. 
The predicted phases can be detected by 
using two already widely employed techniques. First, the removal 
of the confining potentials, and the subsequent presence or absence of 
interferences in the time of flight image, would distinguish 
between phase-coherent and incoherent phases. Second, by ramping-up abruptly 
the lattice potential, it is possible to freeze the spatial density 
correlations, which could be later on probed by means of Bragg scattering. 
The latter should allow to distinguish between homogeneous and modulated 
phases. An independent Bragg analysis for fermions and bosons 
should reveal the formation of composite fermions. 

In this Letter we have shown that the phase diagram for Bose-Fermi lattice gases
in the strong coupling limit is enormously rich, and
contains several novel types of quantum phases involving composite
fermions, which for attractive (repulsive) Bose-Fermi interactions 
are formed by a fermion and one or several bosons (bosonic holes). 
The predicted ground-state solutions include 
delocalized phases (metallic, superfluid), and localized 
ones (density wave and domain insulator). The remarkable development  
of the experimental techniques for cold atomic gases allows
not only for the observability of the predicted phases, but 
also for an unprecedented degree of control not  
available in other condensed matter systems.

In the 1D case, due to the leading role of fluctuations, 
mean-field theories become inaccurate. In
1D we can, however, use the Wigner-Jordan transformation to convert
the effective model into the quantum spin $1/2$ chain, the so-called 
XXZ model \cite{Sachdev}, with a fixed 
magnetization $m_z=\rho_F-1/2$, whose ground state is 
known exactly from Bethe Ansatz \cite{Bethe,Johnson}.
The coefficient characterizing the spin
coupling on the $x-y$ plane will then be $J_{eff}/2$, whereas that in the
$z$ direction will be $K_{eff}/4$. 

Finally, we would like to stress that interesting physics is also
expected when $J$ is comparable to $U$ and $V$. For finite
$J$ the phase diagrams should be extended to three-dimensions by
adding the $J/V$ axis, and will develop a lobe structure in the $J/V-\bar
\mu$ plane, similar to that occurring in MI phases in the Bose-Hubbard
model \cite {Sachdev}. This analysis, as well as the studies of the 
excitations in this system, will be the subject of a separate paper. 

We thank B. Damski, F. E{\ss}ler, H. J. Everts 
and J. Zakrzewski for fruitful discussions.
We acknowledge support from the Deutsche
Forschungsgemeinschaft SFB 407 and SPP1116, the RTN Cold Quantum Gases, 
IST Program EQUIP, ESF PESC BEC2000+, Russian Foundation for Basic 
Research, and the Alexander von Humboldt Foundation.

\end{document}